\documentclass[%
 reprint,
superscriptaddress,
nofootinbib,
 amsmath,amssymb,
 aps,
]{revtex4-2}

\usepackage{graphics,epsfig,psfrag}
\usepackage{color}
\usepackage{graphicx}
\usepackage{dcolumn}
\usepackage{bm}
\usepackage{hyperref}
\usepackage{xcolor}
\usepackage{comment}
\usepackage{enumitem}

\hypersetup{
           breaklinks=true,   
           colorlinks=false,   
           pdfusetitle=true,  
        }

\begin{document}

\title{Deflection of light rays in a moving  medium around a spherically symmetric gravitating object}

\author{Barbora Bezd\v{e}kov\'{a}}
\email{barbora.bezdekova@mff.cuni.cz}
\affiliation
{Institute of Theoretical Physics, Faculty of Mathematics and Physics, Charles University, 18000 Prague, Czech Republic}

\author{Oleg Yu. Tsupko}
\email{tsupkooleg@gmail.com}
\affiliation{ZARM, University of Bremen, 28359 Bremen, Germany}

\author{Christian Pfeifer}
\email{christian.pfeifer@zarm.uni-bremen.de}
\affiliation{ZARM, University of Bremen, 28359 Bremen, Germany}

\date{\today}

\begin{abstract}
In most analytical studies of light ray propagation in curved spacetimes around a gravitating object surrounded by a medium, it is assumed that the medium is a cold nonmagnetized plasma. The distinctive feature of this environment is that the equations of motion of the rays are independent of the plasma velocity, which, however, is not the case in other media. In this paper, we consider the deflection of light rays propagating near a spherically symmetric gravitating object in a moving dispersive medium given by a general refractive index. The deflection is studied when the motion of the medium is defined either as a radially falling onto a gravitating object (e.g., black hole), or rotating in the equatorial plane. For both cases the deflection angles are obtained. These examples demonstrate that fully analytic expressions can be obtained if the Hamiltonian for the rays takes a rather general form as a polynomial in a given momentum component. The general expressions are further applied to three specific choices of refractive indices and these cases are compared. Furthermore, the light rays propagating around a gravitating object surrounded by a generally moving medium are further studied as a small perturbation of the cold plasma model. The deflection angle formula is hence expressed as a sum of zeroth and first order components, where the zeroth order term corresponds to the known cold plasma case and the first order correction can be interpreted as caused by small difference in the refractive index compared to the cold plasma. The results presented in this paper allow to describe the effects caused by the motion of a medium and thus go beyond cold nonmagnetized plasma model.
\end{abstract}

\maketitle

\section{Introduction}
\label{sec:introduction}

Light bending in a gravitational field is one of the famous effects of general relativity. Soon after the theoretical prediction, it has been probed during solar eclipse \cite{Dyson-Eddington-1920, Will-2015-eclipse, Cervantes-Cota-2019-review, Goldoni-2020-eclipse}. The deflection of light by gravity manifested itself in various gravitational lensing phenomena which were also confirmed by observations, see, e.g., \cite{GL1, Blandford-Narayan-1992-review, Wambsganss-1998-review, GL2, Bartelmann-2010-review, Dodelson-GL, Congdon-Keeton-book-2018}.

Light trajectories are affected not only by the gravitational field of a compact object they are passing by, but also by the interaction with a medium through which they propagate. Hence, in calculations focusing on a description of relevant astrophysical situations, it is necessary to take into account that the light propagation is determined by a composition of several competing factors, such as gravitation, refraction, and dispersion.

For an analytical description of light trajectories near gravitating objects in a dispersive and refractive medium, an approach developed many decades ago by Synge \cite{Synge-1960} has recently been applied frequently. This formalism is based on a Hamiltonian description of light in a geometrical optics limit, and it is hence possible to study trajectories of light rays. Investigation of the geometrical optics in a medium in the presence of gravity was presented in Bi\v{c}\'{a}k and Hadrava \cite{Bicak-Hadrava-1975}; see also \cite{Kichenassamy-Krikorian-1985, Kulsrud-Loeb-1992, Krikorian-1999}. Furthermore, light ray propagation in plasma was also deeply studied in a monograph written by Perlick \cite{Perlick-2000}. There, among others, an exact formula for the integral form of light deflection angle in the equatorial plane of a black hole described by the Kerr metric surrounded by cold plasma was derived for the first time. Magnetized plasma was considered by Breuer and Ehlers \cite{Breuer-Ehlers-1980, Breuer-Ehlers-1981, Breuer-Ehlers-1981-AA}, and Broderick and Blandford \cite{Broderick-Blandford-2003, Broderick-Blandford-2003-ASS, Broderick-Blandford-2004}. Gravitational lensing in the presence of refraction due to non-homogeneous plasma was also considered in a weak deflection approximation in earlier works of Muhleman, Ekers and Fomalont \cite{Muhleman-1970} (see also \cite{Muhleman-1966}) and Bliokh and Minakov \cite{Bliokh-Minakov-1989}.

In recent works based on the Synge's formalism \cite{BK-Tsupko-2009, BK-Tsupko-2010}, the light deflection in the presence of plasma was studied first in the weak deflection approximation. These studies assumed a Schwarzschild black hole and discussed both homogeneous and non-homogeneous plasma. In particular, it was shown that the plasma effect is present already in the homogeneous case. Lensing in the weak deflection regime was further considered in the Kerr metric and homogeneous plasma by Morozova et al \cite{Morozova-2013}. The strong deflection regime of gravitational lensing by a Schwarzschild black hole was then studied in Tsupko and Bisnovatyi-Kogan \cite{Tsupko-BK-2013} for a system, where a homogeneous cold plasma was assumed; properties of higher-order images were calculated analytically. Calculation of a black hole shadow in an arbitrary spherically symmetric spacetime in the presence of a cold plasma was presented in Perlick et al \cite{Perlick-Tsupko-BK-2015}. Astrophysical effects of strong light bending near compact objects in the presence of plasma were considered in detail in a series of works by Rogers \cite{Rogers-2015, Rogers-2017-MG14, Rogers-2017a, Rogers-2017b}.

Light propagation in the Kerr spacetime in the presence of a cold nonmagnetized plasma was studied in detail in Perlick and Tsupko \cite{Perlick-Tsupko-2017, Perlick-Tsupko-2024}. The necessary and sufficient condition on the plasma electron density that guarantees the separability of the Hamilton-Jacobi equation for light rays was found. Calculation of a black hole shadow, the deflection angle and different types of orbits were considered. Crisnejo et al \cite{Crisnejo-Gallo-Jusufi-2019} found the orbit equation for light ray propagating in the equatorial plane of stationary and axisymmetric spacetime surrounded by a cold nonmagnetized plasma. Higher-order corrections for the deflection angle of light rays in a non-homogeneous plasma in the weak deflection case were calculated. Separability of the Hamilton-Jacobi equation and shadow for light propagation in a plasma in an axially symmetric and stationary spacetime were studied in \cite{Bezdekova-2022}. General conditions for both the spacetime and plasma which have to be met in order to define the Carter constant were introduced. Light deflection in the equatorial plane of general axially symmetric stationary spacetimes and a dispersive medium of a general refractive index was considered in Bezd{\v{e}}kov{\'a} and Bi{\v{c}}{\'a}k \cite{Bezdekova-2023}, with main application to a cold plasma and the Hartle-Thorne metric, and also to other axisymmetric spacetimes with a quadrupole moment. The deflection angle for light rays in such system was calculated \cite{Bezdekova-2023}.

Among other works about the subject, we refer to a series of works by Crisnejo, Gallo et al. \cite{Crisnejo-Gallo-2018, Crisnejo-Gallo-Rogers-2019, Crisnejo-Gallo-Villanueva-2019, Crisnejo-Gallo-Jusufi-2019, Crisnejo-Gallo-2023-perturbative}, Er and Mao \cite{Er-Mao-2014, Er-Mao-2022}, Kimpson, Wu and Zane \cite{Kimpson-2019a, Kimpson-2019b}. Calculation of black hole shadow in the presence of plasma can also be found in \cite{Perlick-Tsupko-BK-2015, Perlick-Tsupko-2017, Perlick-Tsupko-2024, Perlick-Tsupko-2022, Yan-2018, Chowdhuri-2021-shadow-expand, Badia-Eiroa-2021, Badia-Eiroa-2023, Li-2022, Bezdekova-2022, Briozzo-Gallo-Madler-2023, Huang-2018, Zhang-2023, Kobialko-2024}. For different recent studies of gravitational lensing in the presence of plasma, see further 
\cite{Schulze-Koops-Perlick-Schwarz-2017, Sareny-2019, Turyshev-2019a, Turyshev-2019b, Wagner-Er-2020, Tsupko-BK-2020-microlensing, Sun-Er-Tsupko-2023, Er-Yang-Rogers-2020, Matsuno-2021, Guerrieri-Novello-2022, Chainakun-2022, Kumar-Beniamini-2023, BK-Tsupko-2023-time-delay, Briozzo-Gallo-2023, Kumar-Beniamini-2023, BK-Tsupko-2023-time-delay}.

As discussed above, most of the studies focusing on the light propagation around a gravitating object surrounded by a medium with dispersive and refractive properties assume the cold nonmagnetized plasma model. This is the simplest plasma approximation with interesting properties and astrophysical applications.

The deflection of light in a spherically symmetric spacetime filled by a general dispersive medium was considered by Tsupko \cite{Tsupko-2021}. These results can be used for arbitrary more complicated medium definitions. The calculations were performed only when the medium was assumed to be static.\\

In this paper, we investigate the deflection of light in a moving medium in a spherically symmetric spacetime. It is known that in the simplest case of a cold nonmagnetized plasma, the motion of plasma does not affect the light deflection (see, e.g., \cite{Perlick-2000, Perlick-Tsupko-BK-2015, Schulze-Koops-Perlick-Schwarz-2017} and Section \ref{sec:cold-plasma} for details). However, it is not so in media of a different refractive index. Therefore, the dependence of the deflection angle on the velocity can characterize the difference of another plasma medium from the cold plasma model.

\begin{figure*}
\begin{center}
\includegraphics[width=0.99\textwidth]{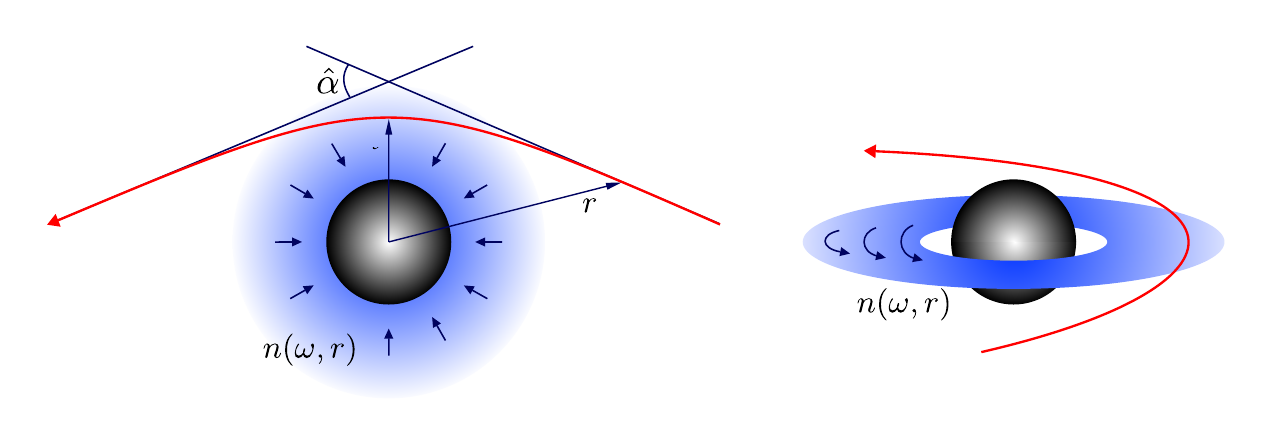}
\end{center}
\caption{Two physical scenarios of a medium motion considered in this paper. Left panel: Spherically symmetric accretion of a dispersive medium with refractive index $n(\omega,r)$ onto a central object. The radial component of velocity $v^r$ is a function of the radial coordinate $r$ only. We calculate the deflection angle $\hat{\alpha}$ of a ray coming from infinity, reaching a minimum radius $R$ and going back to infinity. This case is considered in Sections \ref{sec:radial-falling} and \ref{sec:three-examples}. Right panel: Rotating accretion disk of a dispersive medium in the equatorial plane. The angular component of velocity $v^\varphi$ is a function of the radial coordinate $r$ only. This case is considered in Section \ref{sec:phi-motion}.}
\label{fig:two-models}
\end{figure*}

The paper is organized as follows. We start from the review of the Synge's method and properties of cold plasma model in its context (Section \ref{sec:cold-plasma}). Then, we consider two physically motivated scenarios (Fig.~\ref{fig:two-models}). The first is a spherically symmetric accretion, where the matter is radially falling onto a gravitating object. This is discussed in Section \ref{sec:radial-falling}. Moreover, particular cases of three different refractive indices definitions are compared with each other in Section \ref{sec:three-examples}. The second scenario is a geometrically thin rotating accretion disk in the equatorial plane, where the matter is on circular orbits around a gravitating object, and it is studied in Section \ref{sec:phi-motion}. One more approach is presented in Section \ref{sec:general}, where the effect of a given medium is analyzed as a small perturbation from the cold plasma case and the light deflection is hence presented as a sum of zeroth and first order components. This strategy was also employed in a different context, for example when one studies the propagation of light in effective quantum spacetime models \cite{Addazi:2021xuf}, in cosmological homogeneous and isotropic symmetry \cite{Pfeifer:2018pty, Amelino-Camelia:2023srg}, or in spherical symmetry \cite{Barcaroli:2017gvg, Laanemets-Hohmann-Pfeifer-2022}. Our Conclusions are formulated in Section \ref{sec:conclusions}.

The metric signature is chosen as $\{-,+,+,+\}$. With Latin indices we take $i, k = 0,1,2,3$, resp. $(t,r,\vartheta, \varphi)$, while for Greek indices it is assumed $\alpha, \beta = 1,2,3$, resp. $(r,\vartheta, \varphi)$. It is also used $G=c=1$. Differentiation with respect to the curve parameter present in the Hamiltonian equations is denoted by a dot.

\section{Synge's method and some features of cold plasma case}
\label{sec:cold-plasma}

\subsection{Basic equations of the Synge's method}

In Synge's method \cite{Synge-1960}, geometric optics in an isotropic dispersive medium in the presence of gravity, is based on the application of the so-called medium equation and the resulting Hamiltonian. We emphasize that the method does not cover anisotropic media such as a magnetised plasma.

Let us consider a spacetime filled by a transparent dispersive medium. The spacetime geometry is given by the components of the metric $g^{ik}$. The medium is specified by its refractive index $n$ (which is the reciprocal of the phase speed) and the medium's four-velocity $V^i$. The refractive index is a given function of coordinates $x^i$ and frequency $\omega$:
\begin{equation} \label{n-function} 
n = n(x^i, \omega) \, .
\end{equation}
The medium's four-velocity is a given function of coordinates $x^i$. 
The photon frequency
\begin{equation}
\omega = \omega(p_i, x^i) \, 
\end{equation}
is measured in the instanteneous rest frame of the medium. We have the following relation:
\begin{equation} \label{synge-freq-general}
\omega(p_i, x^i) = - \, p_i V^i  \, .
\end{equation}

Then, the medium equation is \cite{Synge-1960}:
\begin{equation} \label{medium-equation}
n^2 = 1 + \frac{p_i p^i}{\left(p_k V^k \right)^2} \, .
\end{equation}

To apply the Hamiltonian formalism, the equation (\ref{medium-equation}) is rewritten in the form \cite{Synge-1960}
\begin{equation} \label{H-null}
\mathcal{H}(x^i,p_i) = 0 \, ,   
\end{equation}
where the Hamiltonian takes the form
\begin{equation} \label{Hamiltonian-1}
\mathcal{H}(x^i,p_i) = \frac{1}{2} \left\{ g^{ik} p_i p_k - [n^2(x^i, \omega(p_i,x^i)) -1] \left(p_j V^j \right)^2 \right\}  \, .
\end{equation}
Then, the propagation of the light rays is described by the Hamilton’s equations:
\begin{equation} \label{Ham-equations}
\dot{x}^i = \frac{\partial \mathcal{H}}{\partial p_i}  \, , \; \; \dot{p}_i  = - \frac{\partial \mathcal{H}}{\partial x^i} \, .
\end{equation}
We note that the solutions to Hamilton's equations \eqref{Ham-equations} with \eqref{H-null} remain unchanged (up to reparametrisation) if the Hamiltonian (\ref{Hamiltonian-1}) is multiplied with a nowhere vanishing function that may depend on $x^i$ and $p_i$.

It should be emphasized that the Eq.~\eqref{medium-equation} is not the definition of refractive index $n$. To use the method for a specific medium, one should write the refractive index~(\ref{n-function}) as an explicit function of $x^i$ and $\omega$ and substitute it into Eq.~(\ref{medium-equation}) or Eq.~(\ref{Hamiltonian-1}). If medium is non-dispersive, the refractive index is the function of $x^i$ only; in dispersive medium it depends also on the photon frequency (\ref{synge-freq-general}).

\subsection{Distinctive features of cold nonmagnetized plasma}
The important example of a medium is the cold nonmagnetized plasma. First, it is the most relevant in terms of astrophysical applications. Second, the cold plasma provides interesting dispersive properties due to the specific form of a refractive index. The refractive index of a cold plasma is given by
\begin{equation} \label{refr-index-plasma-gen}
n^2(x^i, \omega(p_i, x^i)) = 1 - \frac{\omega_p^2(x^i)}{\omega^2(p_i,x^i)} \, ,
\end{equation}
where $\omega_p$ is the electron plasma frequency, related to electron number density as
\begin{equation}
\omega_p^2(x^i) = \frac{4 \pi e^2}{m_e} N_e(x^i) \, .
\end{equation} 
Substitution of $n$ in the form (\ref{refr-index-plasma-gen}) into the medium equation (\ref{medium-equation}) leads to:
\begin{equation}
- \, \omega_p^2(x^i) = p_i p^i \, .
\end{equation}
The substitution of (\ref{refr-index-plasma-gen}) into Hamiltonian (\ref{Hamiltonian-1}) yields:
\begin{equation} \label{Hamiltonian-2}
\mathcal{H}(x^i,p_i) = \frac{1}{2} \left\{ g^{ik} p_i p_k + \omega_p^2(x^i)  \right\}  \, .
\end{equation}

Already at this stage, we can point out two remarkable properties of light propagation in cold plasma (\ref{refr-index-plasma-gen}).

(i) First, since $\omega(p_i,x^i)$ is dropped out, the Hamiltonian (\ref{Hamiltonian-2}) does not depend on $V^i$. This means that the motion of light rays in cold plasma is independent of the plasma motion. Saying more specific, the light trajectory will be the same for static and moving plasma as far as the constants of motion of the ray are the same. For example, two light rays, one in static plasma and one in moving plasma, will follow the same paths if the photon frequency at infinity and impact parameters are the same for these light rays.

(ii) If cold plasma is homogeneous, i.e., $\omega_p(x^i) = \mbox{const}$, then the substitution of $\omega_p$ by $m$ leads to the Hamiltonian in the form
\begin{equation} \label{Hamiltonian-3}
\mathcal{H}(x^i,p_i) = \frac{1}{2} \left\{ g^{ik} p_i p_k + m^2  \right\}  \, .
\end{equation}
This Hamiltonian describes the motion of massive (as opposite to massless particles, like photon) particles in the gravitational field in vacuum, e.g., \cite{MTW-1973}. Therefore, the motion of light rays in homogeneous plasma is the same as the motion of massive particles (with their mass equal to the plasma frequency) in vacuum. This property was found by Kulsrud and Loeb \cite{Kulsrud-Loeb-1992}, see also further discussion in \cite{Broderick-Blandford-2003} and \cite{BK-Tsupko-2010}. More generally, for \textit{any} $\omega_p$ with no zeros the light rays are timelike geodesics of the conformally rescaled metric $\omega_p^2 g_{ij}$. This more general result is known since
Breuer and Ehlers \cite{Breuer-Ehlers-1981}.

\subsection{Comments on the choice of an observer}
Due to the fact that in the case of a cold plasma its motion does not affect the deflection of rays, it becomes unnecessary to use an observer that is comoving with this plasma. This makes it easier to deal with some of the issues discussed below.

We will consider static and spherically symmetric spacetimes. Let us first write the relation between the photon frequency and the observer's motion.
If the observer is moving with the four-velocity $v^i$, then this observer measures the frequency
\begin{equation} \label{eq:omega-10}
\omega(p_i,x^i) = - p_i v^i   \, .
\end{equation}
This formula is valid for any observer. (It should not be confused with formula (\ref{synge-freq-general}), where the specific observer with $v^i=V^i$ is chosen.) In particular, if the observer is static, then:
\begin{equation} \label{eq:omega-11}
\omega(p_i,x^i) = - p_0 v^0   \, .
\end{equation}
If the observer is radially moving, then there will be also component $v^r$ on the right hand side, i.e.,
\begin{equation} \label{eq:omega-12}
\omega(p_i,x^i) = - p_0 v^0 - p_r v^r   \, .
\end{equation}

Let us now consider a moving cold plasma. For cold plasma, the photon frequency $\omega$ is not present in the Hamiltonian \eqref{Hamiltonian-2}, there is only plasma frequency $\omega_p(r)$. Since the velocity of the observer is only inside $\omega$, we can choose the observer we prefer more. It is convenient to consider a static observer because the formula \eqref{eq:omega-11} is simpler than \eqref{eq:omega-12}. In the papers \cite{Perlick-Tsupko-BK-2015, Perlick-Tsupko-2017, Perlick-Tsupko-2024} about light propagation in plasma we used a static observer. At the same time, all formulas are valid both for static and moving plasma.

In the case of the motion of an arbitrary medium, we cannot choose an arbitrary observer. Synge's approach assumes that we use the observer who is comoving with the medium. 

So the observes's velocity $v^i$ is equal to the medium velocity $V^i$. Correspondingly, in the formula for the photon frequency \eqref{synge-freq-general} we write $V^i$, not $v^i$. Therefore, the right hand side of equation for frequency will depend on the motion of medium. If both medium and observer are static, then we can write simply
\begin{equation} \label{eq:omega-13}
\omega(p_i,x^i) = - p_0 V^0   \, .
\end{equation}
This approach was used by Tsupko in \cite{Tsupko-2021}. If the medium and observer are radially moving, then
\begin{equation} \label{eq:omega-14}
\omega(p_i,x^i) = - p_0 V^0 - p_r V^r   \, .
\end{equation}
This is what we consider in the present paper in the next two Sections.

To conclude, in the case of a cold plasma, it is not necessary to consider static plasma to use the simple equation \eqref{eq:omega-11}. It is possible to use the expression \eqref{eq:omega-11} for a static observer for a moving plasma as well. However, if the medium is different, the use of the simpler expression \eqref{eq:omega-13} is possible only for a static medium (and, consequently, for a static observer, because the observer is comoving with the medium).

\section{Deflection angle of light ray in a radially falling medium}
\label{sec:radial-falling}

In this Section, we will consider the motion of light rays in a spherically symmetric spacetime in the presence of a spherically symmetric medium with non-zero radial velocity (left panel of Fig.~\ref{fig:two-models}). Mainly, we are interested in the case of a radially falling matter onto the central object.

Let us consider a spherically symmetric and static spacetime of a general form
\begin{equation} \label{metric}
ds^2 = - A(r) \, dt^2 + B(r) \, dr^2 + D(r) ( d\vartheta^2 + \sin^2\vartheta \, d\varphi^2) \,  ,
\end{equation}
where $A(r)$, $B(r)$ and $D(r)$ are positive coefficients.
Since we are going to deal with the notion of deflection angle, the condition of asymptotic flatness is required and hence
\begin{equation} \label{as-flatness}
A(r) \to 1, \quad  B(r) \to 1, \quad \frac{D(r)}{r^2} \to 1  
\end{equation}
for $r \to \infty$. We assume that the spacetime (\ref{metric}) is filled by a spherically symmetric dispersive medium with given refractive index $n(r, \omega)$.

The restrictions above will hold throughout the paper. In this Section, we additionally assume that the medium is radially falling onto an object (such as black hole). So, we write the components of a medium four-velocity as
\begin{equation}
V^i = (V^0, V^r, 0, 0) \, .
\end{equation}
Further, we assume that $V^r$ is a known function of $r$, say:
\begin{equation}
V^r = f(r) \, .
\end{equation}
Note that for a radially falling medium, $f(r) < 0$.

The component $V^0$ can now be found from the normalization condition
\begin{equation} \label{normalization_condition}
V_i V^i  = - 1 \, ,
\end{equation}
which yields
\begin{equation} \label{V_0}
    V^0(r) = \sqrt{\frac{1+B(r)f^2(r)}{A(r)}} \, .
\end{equation}
For simplification of further formulas, we will write only $V^0(r)$, bearing in mind that the expression (\ref{V_0}) should be substituted everywhere.

For the photon frequency (\ref{synge-freq-general}) we have:
\begin{equation} \label{omega-r}
\omega(p_r,r)   = - p_0 V^0(r) - p_r V^r = - p_0 V^0(r) - p_r f(r) \, .
\end{equation}
Note that if the motion of medium is given only by a function of $r$, so it does not depend on time, $p_0$ is still the constant of motion. This can be seen from the equation for $\dot{p}_0$ derived from (\ref{Ham-equations}).

Recall that we assume that at $r=\infty$ it holds $A(r) \rightarrow 1$, $B(r) \rightarrow 1$. For simplicity, let us also set that $f(r) \rightarrow 0$ at $r=\infty$.
Taking (\ref{omega-r}) at infinity, we will find the relation between $\omega_0$ (the photon frequency at infinity) and the constant of motion $p_0$, namely
\begin{equation}
p_0 = - \omega_0 \, .
\end{equation}
As a result, the expression (\ref{omega-r}) for $\omega(p_r,r)$ becomes
\begin{equation} \label{omega-r-via-omega-0}
\omega(p_r,r)   =  \omega_0 V^0(r) - p_r f(r) \, .
\end{equation}

Without a loss of generality, we will consider the equatorial plane of metric (\ref{metric}), where $\vartheta=\pi/2$ and $p_\vartheta=0$.
Using the metric coefficients (\ref{metric}) and the expression for photon frequency (\ref{omega-r}) in (\ref{Hamiltonian-1}), we find relevant Hamiltonian in the form
\begin{equation} \label{Hamiltonian-4}
\mathcal{H}=\frac{1}{2} \left[ \frac{p_r^2}{B(r)} + \frac{p_{\varphi}^2}{D(r)} - \frac{\omega_0^2}{A(r)} + w(\omega(p_r,r), r) \right] \, .
\end{equation}
Here we have introduced the function
\begin{equation} \label{def-w}
w(\omega(p_r,r),r) = -(n^2-1) \, \omega^2(p_r,r) \, , \; n = n(\omega(p_r,r),r) \, .
\end{equation}
The function $w(\omega(p_r,r),r)$ is convenient because it better characterizes the main properties of the light propagation in a medium in the gravitational field than refractive index $n$ only. In particular, for cold plasma we have $w = \omega_p^2(r)$.

From the equations of motion for $p_0$ and $p_\varphi$, i.e.,
\begin{equation}
\dot{p}_0 = - \frac{\partial\mathcal{H}}{\partial t} = 0  \, ,
\end{equation}
\begin{equation} 
\dot{p}_\varphi = - \frac{\partial\mathcal{H}}{\partial \varphi} = 0 \, ,
\end{equation}
we find that $p_0$ and $p_\varphi$ are indeed the constants of motion.

The equation for $r$ reads
\begin{equation}
\dot{r} = \frac{\partial\mathcal{H}}{\partial p_r}
= \frac{p_r}{B(r)} \, 
+ \, \frac{1}{2}\frac{\partial w}{\partial \omega}\frac{\partial \omega(p_r,r)}{\partial p_r} \, .
\end{equation} 
Using (\ref{omega-r}) and introducing the notation $w_\omega \equiv \partial w/ \partial \omega$ for compactness, we write
\begin{equation} \label{dot-r-gen}
\dot{r} = \frac{p_r}{B(r)} \, - \, \frac{1}{2} w_\omega f(r) \, .
\end{equation}
Note that $w_\omega = w_\omega(p_r, r; \omega_0)$.

The equation for $\varphi$ is
\begin{equation} \label{dot-phi-gen}
\dot{\varphi}=\frac{\partial\mathcal{H}}{\partial p_\varphi}=\frac{p_{\varphi }}{D(r)} \, .     
\end{equation}

For further calculations, it becomes necessary to express $p_r$. In a general case we can assume that a concrete form of $p_r$ can be found from the condition $\mathcal{H}=0$, where $\mathcal{H}$ 
is given by (\ref{Hamiltonian-4}). This has to be consecutively plugged into (\ref{dot-r-gen}), which will allow one to derive a corresponding expression for $d\varphi/dr$. In the most general case, this can be done numerically.

Let us assume that the value $p_r(r; p_\varphi, \omega_0)$ is known. Then, dividing (\ref{dot-phi-gen}) by (\ref{dot-r-gen}) we find:
\begin{equation}
\frac{d\varphi}{dr} = \frac{p_{\varphi }}{D(r)}
\left[ \frac{p_r(r; p_\varphi, \omega_0)}{B(r)} \, - \, \frac{1}{2} w_\omega(p_r,r; \omega_0) f(r) \right]^{-1} \, .
\end{equation}
In the calculation of the deflection angle we will integrate from the distance of the closest approach $R$ to infinity (see the next Section). In order to have the final angle as a function of $R$ and $\omega_0$ only, we need to exclude $p_\varphi$. The value $p_\varphi(R; \omega_0)$ can be found from the equation
\begin{equation}
\frac{p_r(R; p_\varphi, \omega_0)}{B(R)} \, - \, \frac{1}{2} w_\omega^{(R)} f(R) = 0 \, ,
\end{equation}
where $w_\omega^{(R)} = w_\omega(p_r(R; p_\varphi, \omega_0), R; \omega_0)$.
This condition stems from the fact that $dr/d\varphi=0$ at $r=R$.

\section{Analytical calculation of deflection angle with three particular forms of refractive index}
\label{sec:three-examples}

As follows from the previous Section, explicit analytical formula for $d\varphi/dr$ can be found when it becomes possible to express $p_r$ explicitly from $\mathcal{H}=0$, with $\mathcal{H}$ given by Eq.~(\ref{Hamiltonian-4}). Obviously, it is not possible for a general case because the function $w(\omega(p_r,r),r)$ given by Eq.~(\ref{def-w}) may depend on the photon frequency $\omega(p_r,r)$ in a complicated way. We recall that variable $p_r$ lies inside $\omega(p_r,r)$ through Eq.~(\ref{omega-r-via-omega-0}).

However, a fully analytical solution is possible, for example, with a rather general form of refractive index
\begin{equation} \label{eq:refr-a0-a1-a2}
n^2(r,\omega) = a_0(r) + \frac{a_1(r)}{\omega} + \frac{a_2(r)}{\omega^2} \, ,
\end{equation}
where coefficients $a_0(r)$, $a_1(r)$, and $a_2(r)$ are arbitrary functions of $r$ and $\omega=\omega(p_r,r)$. All these coefficients can be positive, zero or negative, with only restrictions that the whole expression for $n^2$ is non-negative and also that the group velocity in such medium is less than $c$.

With the refractive index (\ref{eq:refr-a0-a1-a2}), the Hamiltonian (\ref{Hamiltonian-4}) is
\begin{gather} \label{Hamiltonian-polyn-1}
\mathcal{H}=\frac{1}{2} \left[ \frac{p_r^2}{B(r)} + \frac{p_{\varphi}^2}{D(r)}  - \frac{\omega_0^2}{A(r)} \right.\\
\left. - \, a_0(r) \, \omega^2(p_r,r) - a_1(r) \, \omega(p_r,r) - a_2(r) + \omega^2(p_r,r) \right] \, , \nonumber
\end{gather}
where $\omega(p_r,r)$ is given by Eq.~(\ref{omega-r-via-omega-0}).

The Hamiltonian (\ref{Hamiltonian-polyn-1}) turns out to be a quadratic polynomial of $p_r$, which makes it possible to explicitly express $p_r$ and, accordingly, obtain the deflection angle in a compact form. Let us rewrite (\ref{Hamiltonian-polyn-1}) as
\begin{equation} \label{eq:Hamilt-ABC}
\mathcal{H}= \frac{1}{2} \left[ \mathcal{A}_r(r) \, p_r^2 + 2\mathcal{B}_r(r) \, p_r + \mathcal{C}_r(r,p_\varphi) \right],
\end{equation}
where $\mathcal{A}_r(r)$, $\mathcal{B}_r(r)$, $\mathcal{C}_r(r,p_\varphi)$ are some functions of $r$ (which should not be confused with the metric coefficients of (\ref{metric})). Note that they also include $\omega_0$ as a parameter. For our choice of the medium velocity, we can also set
\begin{equation}
\mathcal{C}_r(r,p_\varphi) = \mathcal{C}_{r1}(r) + \frac{p_\varphi^2}{D(r)} - \frac{\omega_0^2}{A(r)} \, .
\end{equation}
The coefficients in the Hamiltonian (\ref{eq:Hamilt-ABC}) are related with the metric coefficients and functions in the refractive index (\ref{eq:refr-a0-a1-a2}) as
\begin{equation}
    \mathcal{A}_r(r) = \frac{1}{B(r)} + (1-a_0(r)) f^2(r) \, ,
\end{equation}
\begin{equation}
    \mathcal{B}_r(r) = \left[ (a_0(r)-1) \, \omega_0 V^0(r) + \frac{1}{2}a_1(r) \right] f(r) \, ,
\end{equation}
\begin{gather}
\mathcal{C}_{r1}(r) =  - \, a_2(r) \\
    + \, \omega_0 V^0(r)\left[(1-a_0(r)) \, \omega_0 V^0(r)-a_1(r)\right] \, . \nonumber
\end{gather}

Let us now perform general calculations for the Hamiltonian in the form (\ref{eq:Hamilt-ABC}).

The corresponding equations of motion then read simply as
\begin{equation} 
\dot{r} =  \mathcal{A}_r(r) \, p_r + \mathcal{B}_r(r) \, ,
\end{equation}
and
\begin{equation}  \label{eq:dot-phi-equation}
\dot{\varphi} = \frac{p_\varphi}{D(r)} \, .
\end{equation}

In a full generality, condition $\mathcal{H}=0$ returns expression
\begin{equation}
    p_r = \frac{-\mathcal{B}_r(r) \pm\sqrt{\mathcal{B}_r^2(r)-\mathcal{A}_r(r) \, \mathcal{C}_r(r,p_\varphi)}}{\mathcal{A}_r(r)} \, ,
\end{equation}
or
\begin{equation}
\mathcal{A}_r(r) \, p_r + \mathcal{B}_r(r) = \pm\sqrt{\mathcal{B}_r^2(r)-\mathcal{A}_r(r) \, \mathcal{C}_r(r,p_\varphi)} \, ,
\end{equation}
which starts to be convenient after plugging into formula for $\dot{r}$ because it then yields
\begin{equation}
    \dot{r} = \pm\sqrt{\mathcal{B}_r^2(r)-\mathcal{A}_r(r) \, \mathcal{C}_r(r,p_\varphi)} \, .
\end{equation}
This significantly simplifies the calculations, as we are now dealing with a basic algebraic expression, and the orbit equation can be written as
\begin{equation}
\frac{d\varphi}{dr} = \pm\frac{p_\varphi}{D(r)} \left[\mathcal{B}_r^2(r) - \mathcal{A}_r(r) \, \mathcal{C}_r(r,p_\varphi) \right]^{-1/2} \, .
\end{equation}

This expression can be further abbreviated as
\begin{equation}
    \frac{d\varphi}{dr}=\pm\frac{1}{\sqrt{\mathcal{A}_r(r)D(r)}}\left(\frac{\omega_0^2}{p_\varphi^2}h^2(r) - 1\right)^{-1/2} \, ,
\end{equation}
where we introduced the function
\begin{equation}\label{h_radially}
    h^2(r)=\frac{D(r)}{A(r)}\left[1+\frac{A(r)}{\omega_0^2}\left(\frac{\mathcal{B}_r^2(r)}{\mathcal{A}_r(r)}-\mathcal{C}_{r1}(r)\right)\right] \, .
\end{equation}
Applying the condition at $r=R$, one finds out that
\begin{equation}
\frac{p_\varphi^2}{\omega_0^2}=h^2(R) \, ,
\end{equation}
and hence
\begin{equation} \label{eq:traj-three-ex}
\frac{d\varphi}{dr} = \pm \frac{1}{\sqrt{\mathcal{A}_r(r)D(r)}}\left(\frac{h^2(r)}{h^2(R)} - 1\right)^{-1/2} \, .
\end{equation}

Let us calculate the deflection angle $\hat{\alpha}$ of light ray moving from infinity to a gravitating object, reaching the closest approach value $R$ and moving again to infinity (Fig.~\ref{fig:two-models}). For a convenience, we can assume that the light ray moves in such a way that its $\varphi$-coordinate increases ($d\varphi>0$); by Eq.~\eqref{eq:dot-phi-equation}, we have also $p_\varphi>0$ for this ray. Then, if the ray approaches the center ($r$-coordinate decreases, $dr<0$), we should use minus-sign in the equation \eqref{eq:traj-three-ex}. And vice versa, if the ray moves away from the center ($r$-coordinate increases, $dr>0$), the plus-sign in the equation \eqref{eq:traj-three-ex} should be chosen.

The change of $\varphi$-coordinate for a considered ray is given by
\begin{gather}
\Delta \varphi
= \int \limits_\infty^R \frac{d\varphi}{dr} 
\, dr + \int \limits_R^\infty \frac{d\varphi}{dr} \, dr = \nonumber \\
= - \int \limits_\infty^R  \frac{1}{\sqrt{\mathcal{A}_r(r)D(r)}}\left(\frac{h^2(r)}{h^2(R)} - 1\right)^{-1/2} + \nonumber \\
+ \int \limits_R^\infty  \frac{1}{\sqrt{\mathcal{A}_r(r)D(r)}}\left(\frac{h^2(r)}{h^2(R)} - 1\right)^{-1/2} = \nonumber \\
= 2 \int \limits_R^\infty  \frac{1}{\sqrt{\mathcal{A}_r(r)D(r)}}\left(\frac{h^2(r)}{h^2(R)} - 1\right)^{-1/2} \, .
\end{gather}
The value of $\Delta \varphi$ is positive. Taking into account that the straight-line propagation of light corresponds to $\Delta \varphi=\pi$, we find for the deflection angle $\hat{\alpha}$:
\begin{equation} 
\hat{\alpha} =  \Delta \varphi - \pi \, ,
\end{equation}
or
\begin{equation} \label{eq:angle-three-ex}
\hat{\alpha} =  2 \int \limits_R^\infty \frac{1}{\sqrt{\mathcal{A}_r(r)D(r)}}\left(\frac{h^2(r)}{h^2(R)} - 1\right)^{-1/2} dr \, - \, \pi \, .
\end{equation}
Since we got rid of constant $p_\varphi$, the formula (\ref{eq:angle-three-ex}) depends only on the distance of the closest approach $R$ and the frequency of the photon at infinity $\omega_0$. The dependence of $h(r)$ on $\omega_0$ means that an observed lensing picture will look like a ``rainbow''. For cold plasma case, the effects are non-negligible only for radioband, and the vacuum case is recovered if $\omega_0 \to \infty$. Recall that in vacuum the deflection angle depends on $R$ only. For vacuum case, formulas for deflection angle in spherically symmetric metric can be found, e.g., in \cite{weinberg72,virbhadra98}.

As a side remark, such a rainbow feature also emerges in the context of quantum gravity phenomenology, when one models the propagation of light on quantum spacetime by a modified dispersion relation \cite{Glicenstein:2019rzj, Laanemets-Hohmann-Pfeifer-2022}. In the context of quantum gravity the relevant frequency depends on the model, but usually non-negligible effects are assumed to emerge for very high energetic photons \cite{Addazi:2021xuf} since the modification of the general relativistic dispersion relation is usually assumed to be suppressed by the Planck energy. In this case, the classical vacuum result is obtained by sending the Planck energy to infinity.

We remind that the formulas (\ref{eq:traj-three-ex}) and (\ref{eq:angle-three-ex}) were derived under the assumption that for the refractive index (\ref{eq:refr-a0-a1-a2}) the Hamiltonian can be written in the form (\ref{eq:Hamilt-ABC}) with some functions $\mathcal{A}_r(r)$, $\mathcal{B}_r(r)$, and $\mathcal{C}_r(r,p_\varphi)$. This allows us to separate explicitly the dependence on $p_r$ in the Hamiltonian. After defining these functions for a specific form of refractive index, the formula for deflection angle can be easily written from the equation given above.

Let us now specify three examples of the medium.\\

\textit{Example 1.}

Cold plasma model with the refractive index of the form:
\begin{equation} \label{refr-index-plasma-r}
n^2(r, \omega) = 1 - \frac{\omega_p^2(r)}{\omega^2} \, , \quad \omega = \omega(p_r,r) \, .
\end{equation}

In this case, the function $w(\omega(p_r,r),r)$ defined in (\ref{def-w}) reduces to a simple form, independent of $\omega(p_r,r)$:
\begin{equation} 
w(\omega(p_r,r),r) = \omega_p^2(r) \, .
\end{equation}
Correspondingly, the Hamiltonian (\ref{Hamiltonian-4}) reduces to
\begin{equation} \label{Hamiltonian-4a}
\mathcal{H}=\frac{1}{2} \left[ \frac{p_r^2}{B(r)} + \frac{p_{\varphi}^2}{D(r)} - \frac{\omega_0^2}{A(r)} + \omega_p^2(r) \right] \, .
\end{equation}
Then, the individual functions $\mathcal{A}_r(r)$, $\mathcal{B}_r(r)$, and $\mathcal{C}_{r1}(r)$ are as follows:
\begin{equation}
    \mathcal{A}_r(r) = \frac{1}{B(r)} \, ,
\end{equation}
\begin{equation}
    \mathcal{B}_r(r) = 0 \, ,
\end{equation}
\begin{equation}
   \mathcal{C}_{r1}(r) = \omega_p^2(r) \, .
\end{equation}
Deflection angle (\ref{eq:angle-three-ex}) reduces to
\begin{equation} \label{eq:angle-cold-plasma}
\hat{\alpha} =  2 \int \limits_R^\infty \frac{\sqrt{B(r)}}{\sqrt{D(r)}}\left(\frac{h^2(r)}{h^2(R)} - 1\right)^{-1/2} dr \, - \, \pi \, ,
\end{equation}
with the function $h(r)$ of the form
\begin{equation} \label{eq:h-cold-plasma}
    h^2(r)=\frac{D(r)}{A(r)}\left(1-A(r)\frac{\omega_p^2(r)}{\omega_0^2}\right) \, .
\end{equation}
Equation (\ref{eq:angle-cold-plasma}) with (\ref{eq:h-cold-plasma}) was found in Perlick et al. \cite{Perlick-Tsupko-BK-2015}, see Eq.~(20) there.\\

\textit{Example 2.}

Non-dispersive medium, so the refractive index does not depend on the photon frequency $\omega$, but it is the function of $r$ only:
\begin{equation} \label{refr-index-n-r}
n = n(r) \, .
\end{equation}
In this case, the Hamiltonian in the equatorial plane is
\begin{equation}
\mathcal{H} = \frac{1}{2} \left[\frac{p_r^2}{B(r)} + \frac{p_{\varphi}^2}{D(r)} - \frac{\omega_0^2}{A(r)} + w(\omega, r)\right],
\end{equation}
where $n$ is not a function of $\omega$ anymore.

The coefficients $\mathcal{A}_r(r)$, $\mathcal{B}_r(r)$, and $\mathcal{C}_{r1}(r)$ are as follows:
\begin{equation}
    \mathcal{A}_r(r) = \frac{1}{B(r)}-(n^2(r)-1)f^2(r) \, ,
\end{equation}
\begin{equation}
    \mathcal{B}_r(r) = (n^2(r)-1) \, \omega_0V^0(r)f(r) \, ,
\end{equation}
\begin{equation}
     \mathcal{C}_{r1}(r) = -(n^2(r)-1)(\omega_0 V^0(r))^2 \, .
\end{equation}
And the function $h(r)$ takes the form:
\begin{equation}
    h^2(r)=\frac{D(r)}{A(r)}\left(1+A(r)\frac{(n^2(r)-1)(V^0(r))^2}{1-(n^2(r)-1)B(r)f^2(r)}\right) \, .
\end{equation} \\

\textit{Example 3.}

A medium with the refractive index of the form:
\begin{equation} \label{refr-index-third-example}
n^2(r,\omega) = 1 + \frac{a(r)}{\omega}  \, , \quad \omega=\omega(p_r,r) \, .
\end{equation}
In this case, the Hamiltonian in the equatorial plane takes the form
\begin{equation}
\mathcal{H} = \frac{1}{2} \left[ \frac{p_r^2}{B(r)} + \frac{p_{\varphi}^2}{D(r)} - \frac{\omega_0^2}{A(r)} - a(r) \, \omega(p_r,r) \right].
\end{equation}
The coefficients $\mathcal{A}_r(r)$, $\mathcal{B}_r(r)$, and $\mathcal{C}_{r1}(r)$ are as follows:
\begin{equation}
    \mathcal{A}_r(r) = \frac{1}{B(r)} \, ,
\end{equation}
\begin{equation}
    \mathcal{B}_r(r) = \frac{1}{2}a(r)f(r) \, , 
\end{equation}
\begin{equation}
     \mathcal{C}_{r1}(r) = - a(r) \, \omega_0 V^0(r) \, .
\end{equation}
And the function $h(r)$ takes the form:
\begin{equation}
h^2(r) = \frac{D(r)}{A(r)} \left[ 1 + A(r)\frac{a^2(r)}{\omega_0^2} \left(\frac{1}{4}B(r)f^2(r) + \frac{\omega_0 V^0(r)}{a(r)} \right) \right] \, .
\end{equation}

\section{Deflection angle in a case of equatorial rotation of medium}
\label{sec:phi-motion}

In this Section, we will consider another case of a medium motion (right panel of Fig.~\ref{fig:two-models}). We assume that the medium is rotating in the equatorial plane of a compact object (e.g., a black hole) in the form of a geometrically thin accretion disk and consider the light propagation in this plane only. The form of the metric \eqref{metric} and conditions \eqref{as-flatness} remain the same.

In the case under consideration, the medium four-velocity is defined as
\begin{equation}
V^i = (V^0, 0, 0, V^\varphi) \, .
\end{equation}
Similarly to the previous case, we want $V^\varphi$ to be a known function of $r$ which reads
\begin{equation}
V^\varphi = f(r) \, .
\end{equation}
Since the velocity does not depend on $t$, component $p_0$ is still the constant of motion.

As in the previous case, form of $V^0$ is set from the normalization of four-velocity, and in this case one gets
\begin{equation}\label{V-0-phi}
    V^0(r)=\sqrt{\frac{1+D(r)f^2(r)}{A(r)}} \, .
\end{equation}
In further expressions, we will restrict ourselves on writing $V^0(r)$ only, but it can be substituted as the form of (\ref{V-0-phi}) anytime.

The photon frequency (\ref{synge-freq-general}) in this case reduces to
\begin{equation} \label{omega-r-phi}
\omega(p_\varphi,r)   = - p_0 V^0(r) - p_\varphi V^\varphi = - p_0 V^0(r) - p_\varphi f(r) \, .
\end{equation}

Because the used spacetime metric did not change and it is still assumed that $f(r) \rightarrow 0$ at $r=\infty$ , it therefore holds that $p_0 = - \omega_0 $. As a result, the expression (\ref{omega-r-phi}) for $\omega(p_\varphi,r)$ becomes
\begin{equation} \label{omega-r-phi-via-omega-0}
\omega(p_\varphi,r)   =  \omega_0 V^0(r) - p_\varphi f(r) \, .
\end{equation}

We will consider the photon motion in the equatorial plane.
Assuming $\vartheta=\pi/2$, $p_\vartheta=0$, the corresponding Hamiltonian reads
\begin{equation} \label{Hamiltonian-phi}
\mathcal{H}=\frac{1}{2} \left[\frac{p_r^2}{B(r)}+\frac{p_{\varphi}^2}{D(r)}-\frac{\omega_0^2}{A(r)}+ w(\omega(p_\varphi,r), r)\right] \, ,
\end{equation}
which looks the same as (\ref{Hamiltonian-4}), but here the function
\begin{equation}
w(\omega(p_\varphi,r),r) = -(n^2-1) \, \omega^2(p_\varphi, r) , \; n = n(\omega(p_\varphi,r), r) \, ,
\end{equation}
is defined with the usage of Eq.~(\ref{omega-r-phi-via-omega-0}).

Due to a different definition of $\omega$ given by expression (\ref{omega-r-phi-via-omega-0}) rather than expression (\ref{omega-r-via-omega-0}), the Hamiltonian (\ref{Hamiltonian-phi}) contains components $p_r$ and $p_\varphi$ in another way than the Hamiltonian (\ref{Hamiltonian-4}). This relation leads to a difference in the derivation of the deflection angle. Indeed, in the Hamiltonian (\ref{Hamiltonian-4}) we had the momentum component $p_r^2$ in the first term and also inside the term $w(\omega(p_r,r),r)$, while the component $p_\varphi$ was only in the second term. Correspondingly, the component $p_\varphi$ is easily expressed from the equation $\mathcal{H}=0$, but the component $p_r$ not. In the case of the Hamiltonian (\ref{Hamiltonian-phi}), the component $p_r$ is only in the first term, whereas the component $p_\varphi$ is in the second term and also inside the function $w(\omega(p_\varphi,r),r)$. Hence, in this case we can easily express the component $p_r$ from $\mathcal{H}=0$, but not $p_\varphi$.

The equations of motion are
\begin{gather}
\dot{r}=\frac{\partial\mathcal{H}}{\partial p_r}=\frac{p_r}{B(r)} \, ,\\
    \dot{\varphi}=\frac{\partial\mathcal{H}}{\partial p_\varphi}=
     \frac{p_{\varphi}}{D(r)}+\frac{1}{2}\frac{\partial w}{\partial \omega}\frac{\partial \omega(p_\varphi,r)}{\partial p_\varphi} \, .
\end{gather}
Note that the equations of motion for $p_0$ and $p_\varphi$ remain the same as in the previous case (see Section \ref{sec:radial-falling}), and hence one can directly see that $p_0$ and $p_\varphi$ are still the constants of motion. We can further explicitly calculate $\partial \omega(p_\varphi,r)/\partial p_\varphi$. This leads to the expression
\begin{equation}
    \dot{\varphi}=\frac{p_{\varphi}}{D(r)}-\frac{1}{2} w_\omega(r, p_\varphi; \omega_0) f(r) \, .
\end{equation}

In this case, contrary to the case in Section \ref{sec:radial-falling}, it is possible to easily express $p_r$ from the equation $\mathcal{H}=0$. This yields
\begin{equation}
p_r = \pm\sqrt{B(r)}\sqrt{\frac{\omega_0^2}{A(r)}-\frac{p_{\varphi}^2}{D(r)}-w(\omega(p_\varphi,r), r)} \, .
\end{equation}
Plugging the obtained formula into previous equation leads to the following equation of the trajectory:
\begin{equation}
\frac{d\varphi}{dr} =
\pm \frac{\sqrt{B(r)}}{\sqrt{D(r)}} \left\{ \frac{\frac{D(r)}{p_{\varphi}^2} \left[ \frac{\omega_0^2}{A(r)}-w(\omega(p_\varphi,r), r)\right] - 1 }{\left[ 1-\frac{D(r)}{2p_{\varphi}} w_\omega(r, p_\varphi;\omega_0) f(r) \right]^2} \right\}^{-1/2}
 \, .
\end{equation}
To find an exact value of the deflection angle, one has to again get rid of $p_\varphi$ in the final equation and express the formula as a function of $R$ and $\omega_0$ only. In this case, when assuming that $dr/d\varphi=0$ at $r=R$, it is necessary to find form of $p_\varphi(R; \omega_0)$ from the simple condition
\begin{equation}
p_r(R; p_\varphi, \omega_0)= 0 \, . 
\end{equation} \\

Moreover, one can now perform a similar calculation as was done at the beginning of Section \ref{sec:three-examples}, but with respect to $p_\varphi$.
We assume again the general form of the refractive index (\ref{eq:refr-a0-a1-a2}), i.e.,
\begin{equation} \label{refr-index-a0-a1-a2-phi}
n^2(r,\omega) = a_0(r) + \frac{a_1(r)}{\omega} + \frac{a_2(r)}{\omega^2} \, ,
\end{equation}
but here the form of $\omega=\omega(p_\varphi,r)$ is given by (\ref{omega-r-phi-via-omega-0}). The Hamiltonian (\ref{Hamiltonian-phi}) hence takes the form
\begin{gather} \label{Hamiltonian-polyn-phi}
\mathcal{H}=\frac{1}{2} \left[ \frac{p_r^2}{B(r)} + \frac{p_{\varphi}^2}{D(r)}  - \frac{\omega_0^2}{A(r)} \right.\\
\left. - \, a_0(r) \, \omega^2(p_\varphi,r) - a_1(r) \, \omega(p_\varphi,r) - a_2(r) + \omega^2(p_\varphi,r) \right] \, . \nonumber
\end{gather}
It is again identical to \eqref{Hamiltonian-polyn-1} with the only difference in the form of $\omega(p_\varphi,r)$ defined here with \eqref{omega-r-phi-via-omega-0}.

We will rewrite the Hamiltonian \eqref{Hamiltonian-polyn-phi} as
\begin{equation} \label{eq:Hamilt-ABC-phi}
\mathcal{H}= \frac{1}{2} \left[ \mathcal{A}_\varphi(r) \, p_\varphi^2 + 2\mathcal{B}_\varphi(r) \, p_\varphi + \mathcal{C}_\varphi(r,p_r) \right] \, ,
\end{equation}
where $\mathcal{A}_\varphi(r), \mathcal{B}_\varphi(r), \mathcal{C}_\varphi(r,p_r)$ are the functions of $r$ which should not be confused with the functions $\mathcal{A}_r(r), \mathcal{B}_r(r), \mathcal{C}_r(r,p_\varphi)$ defined in the case when $V^r\ne0$ (Section \ref{sec:three-examples}). However, they again include $\omega_0$ as a parameter. For the suggested form of the medium velocity, one can assume
\begin{equation}
\mathcal{C}_\varphi(r,p_r) = \mathcal{C}_{\varphi1}(r) + \frac{p_r^2}{B(r)} \, .
\end{equation}

The functions $\mathcal{A}_\varphi(r), \mathcal{B}_\varphi(r), \mathcal{C}_\varphi(r,p_r)$ for the refractive index (\ref{refr-index-a0-a1-a2-phi}) then take the form
\begin{gather}
\mathcal{A}_\varphi(r)=\frac{1}{D(r)}+(1-a_0(r))f^2(r) \, , \\
\mathcal{B}_\varphi(r)=\left[(a_0(r)-1) \, \omega_0V^0(r)+\frac{1}{2}a_1(r)\right]f(r) \, , \\
\mathcal{C}_\varphi(r,p_r)=\frac{p_r^2}{B(r)}-\frac{\omega_0^2}{A(r)} \\
+ \, \omega_0 V^0(r) \left[ (1-a_0(r)) \, \omega_0 V^0(r)-a_1(r) \right] - a_2(r) \, . \nonumber
\end{gather}
Note that function $\mathcal{B}_\varphi(r)$ is identical with function $\mathcal{B}_r(r)$, while functions $\mathcal{A}_\varphi(r)$, $\mathcal{C}_\varphi(r)$ differ from functions $\mathcal{A}_r(r)$, $\mathcal{C}_r(r)$.

The equations of motion in this notation are
\begin{equation} 
\dot{r} =  \frac{p_r}{B(r)} \, ,
\end{equation}
\begin{equation} \label{eq:dot-phi-eq-02}
\dot{\varphi} = \mathcal{A}_\varphi(r) \, p_\varphi + \mathcal{B}_\varphi(r) \, .
\end{equation}

The condition $\mathcal{H}=0$ in this case just leads to
\begin{equation}
    p_r = \pm \sqrt{B(r) \left[-\mathcal{A}_\varphi(r) \, p_\varphi^2 - 2 \mathcal{B}_\varphi(r) \, p_\varphi - \mathcal{C}_{\varphi1}(r)\right] } \, ,
\end{equation}
and when plugging this into formula for $\dot{r}$, it gives
\begin{equation}
    \dot{r} = \pm\sqrt{\frac{-\mathcal{A}_\varphi(r) \, p_\varphi^2 - 2 \mathcal{B}_\varphi(r) \, p_\varphi - \mathcal{C}_{\varphi1}(r)}{B(r)}} \, .
\end{equation}
This can be applied in the orbit equation which further reads
\begin{gather} \label{orbit-eq-phi-case}
    \frac{d\varphi}{dr} = \pm \sqrt{B(r)\mathcal{A}_\varphi(r)} \left\{ \frac{\mathcal{B}^2_\varphi(r) - \mathcal{A}_\varphi(r) \, \mathcal{C}_{\varphi1}(r)}{(\mathcal{A}_\varphi(r) \, p_\varphi + \mathcal{B}_\varphi(r))^2}-1 \right\}^{-1/2}.
\end{gather}

As usual, it is desirable to get rid of the constant of motion $p_\varphi$. For that, we need to find an expression describing the dependence of $p_\varphi$ on the closest radial distance $R$ and photon frequency at infinity $\omega_0$. In this case, it leads to a quadratic equation with solutions:
\begin{equation} \label{eq:p-phi-plus-minus}
    p_\varphi = \frac{-\mathcal{B}_\varphi(R) \pm \sqrt{\mathcal{B}^2_\varphi(R)-\mathcal{A}_\varphi(R) \, \mathcal{C}_{\varphi1}(R)}}{\mathcal{A}_\varphi(R)} \, .
\end{equation}
This formula can be used for expressing $p_\varphi$ with a concrete choice of the Hamiltonian and refractive index and plugged into the orbit equation (\ref{orbit-eq-phi-case}) . 

To make the deflection angle formula more compact, it is useful to define function $h(r)$ similarly to (\ref{h_radially}) as it was introduced in the previous sections, but in this case it is more general and it reads
\begin{equation}
   h(r)= \frac{1}{\mathcal{A}_\varphi(r)} \sqrt{\mathcal{B}^2_\varphi(r)-\mathcal{A}_\varphi(r) \, \mathcal{C}_{\varphi1}(r) } \, .
\end{equation}

With this notation, Eq.~\eqref{eq:p-phi-plus-minus} looks as
\begin{equation} \label{eq:p-phi-plus-minus-02}
p_\varphi =
- \frac{\mathcal{B}_\varphi(R)}{\mathcal{A}_\varphi(R)} \pm h(R) \, .
\end{equation}
This can also be rewritten as 
\begin{equation} \label{eq:p-phi-plus-minus-03}
\mathcal{A}_\varphi(R) \, p_\varphi + \mathcal{B}_\varphi(R) = \pm \, h(R) \, .
\end{equation}
This formulation is particularly convenient for the given definitions of the Hamiltonian and refractive index and it can be applied in the orbit equation to get
\begin{gather} \label{orbit-equation-phi-case-final}
\frac{d\varphi}{dr} = \pm \sqrt{B(r)\mathcal{A}_\varphi(r)}\\ \times \left[\frac{h^2(r)} {\left(\frac{\mathcal{B}_\varphi(r)}{\mathcal{A}_\varphi(r)}-\frac{\mathcal{B}_\varphi(R)}{\mathcal{A}_\varphi(R)} \pm h(R) \right)^2}-1 \right]^{-1/2} \, . \nonumber
\end{gather}

The orbit equation (\ref{orbit-equation-phi-case-final}) contains plus-minus sign not only in front of all terms in right hand side (as it was in (\ref{eq:traj-three-ex})), but also in front of the term $h(R)$. This additional plus-minus sign arises from the formulas \eqref{eq:p-phi-plus-minus-02} and \eqref{eq:p-phi-plus-minus-03}. It refers to the fact that due to the presence of the medium rotation in $\varphi$-direction, the deflection will be different in the co-rotation and counter-rotation cases. In our consideration, we do not assign a particular sign to quantity $f(r)$, so we will differ these two situations on the basis of increasing or decreasing $\varphi$-coordinate. Recall that in the previous case (Sections \ref{sec:radial-falling} and \ref{sec:three-examples}) we could consider only the case $\dot{\varphi}>0$ without the loss of generality. 

First, let us assume the motion with increasing $\varphi$-coordinate ($d\varphi>0$ or $\dot{\varphi}>0$), same as we did in Section \ref{sec:three-examples} after Eq.~(\ref{eq:traj-three-ex}). Then, by \eqref{eq:dot-phi-eq-02}, the combination $\mathcal{A}_\varphi(r) \, p_\varphi + \mathcal{B}_\varphi(r)$ is positive. It means that the plus sign should be chosen in \eqref{eq:p-phi-plus-minus-02} and \eqref{eq:p-phi-plus-minus-03}. Correspondingly, the plus sign should be used along with $h(R)$ in formula \eqref{orbit-equation-phi-case-final}.

Considering a ray moving from infinity to $R$ and then again to infinity, we choose plus-minus in front of all terms in (\ref{orbit-equation-phi-case-final}) in the same manner as in Section \ref{sec:three-examples} after Eq.~(\ref{eq:traj-three-ex}). As a result, we can calculate $\Delta \varphi$ which will be positive in this case:
\begin{gather} \label{delta-phi-01}
\Delta \varphi = 2 \int \limits_R^\infty \sqrt{B(r)\mathcal{A}_\varphi(r)} \\
\times\left[\frac{h^2(r)} {\left(\frac{\mathcal{B}_\varphi(r)}{\mathcal{A}_\varphi(r)}-\frac{\mathcal{B}_\varphi(R)}{\mathcal{A}_\varphi(R)} + h(R) \right)^2}-1 \right]^{-1/2} dr \, . \nonumber
\end{gather}
Note the plus sign in front of $h(R)$.

Now, let us assume the motion with decreasing $\varphi$-coordinate ($d\varphi<0$, or $\dot{\varphi} <0$). Then, by \eqref{eq:dot-phi-eq-02}, the combination $\mathcal{A}_\varphi(r) \, p_\varphi + \mathcal{B}_\varphi(r)$ is negative. Correspondingly, the minus sign should be chosen along with $h(R)$ in \eqref{eq:p-phi-plus-minus-02}, \eqref{eq:p-phi-plus-minus-03} and \eqref{orbit-equation-phi-case-final}. The value $\Delta \varphi$ will be negative in this case. After an appropriate choice of plus-minus in front of all terms in (\ref{orbit-equation-phi-case-final}) we find:
\begin{gather} \label{delta-phi-02}
\Delta \varphi = - 2 \int \limits_R^\infty \sqrt{B(r)\mathcal{A}_\varphi(r)} \\
\times\left[\frac{h^2(r)} {\left(\frac{\mathcal{B}_\varphi(r)}{\mathcal{A}_\varphi(r)}-\frac{\mathcal{B}_\varphi(R)}{\mathcal{A}_\varphi(R)} - h(R) \right)^2}-1 \right]^{-1/2} dr \, . \nonumber
\end{gather}
Note the minus sign in front of $h(R)$.

The formulas (\ref{delta-phi-01}) and (\ref{delta-phi-02}) can be combined together with the help of deflection angle $\hat{\alpha}$ which is  usually considered as a positive value:
\begin{equation}
\hat{\alpha} = \pm \Delta \varphi - \pi \, .
\end{equation}
We obtain finally the formula for the deflection angle:
\begin{gather} \label{eq:angle-phi-motion}
\hat{\alpha} = 2 \int \limits_R^\infty \sqrt{B(r)\mathcal{A}_\varphi(r)} \\
\times\left[\frac{h^2(r)} {\left(\frac{\mathcal{B}_\varphi(r)}{\mathcal{A}_\varphi(r)}-\frac{\mathcal{B}_\varphi(R)}{\mathcal{A}_\varphi(R)} \pm h(R) \right)^2}-1 \right]^{-1/2} dr - \pi . \nonumber
\end{gather}
Here plus sign corresponds to the motion with $\dot{\varphi}>0$ and minus sign corresponds to the motion with $\dot{\varphi}<0$. Co-rotation and counter-rotation of a light ray with the medium depends not only on the sign of $\dot{\varphi}$, but also on the sign of $V^\varphi$-component given by $f(r)$. It should also be noted that here we assume that $\omega_0$ is positive for future oriented light rays; see the discussion in Section II of \cite{Perlick-Tsupko-2024}.

It is interesting to compare this result (a spherically symmetric compact object with a rotating moving medium) with the Kerr case in cold plasma, see Section VI of Perlick and Tsupko \cite{Perlick-Tsupko-2024} and \cite{Perlick-2000}, and also with a general deflection angle formula for an axially symmetric object presented in \cite{Bezdekova-2023}. \\

\section{Deflection angle in a case of small perturbation of cold plasma case}
\label{sec:general}

In this Section, let us focus on the case when we can linearize the refractive index as
\begin{equation} \label{eq:refr-index-linearized}
n = n_0 + \varepsilon n_1   \, ,
\end{equation}
where $n_0$ is a cold plasma refractive index (\ref{refr-index-plasma-r}), and $\varepsilon \ll 1$ is a small parameter, while $n_1$ is some known function of $\omega$ and $r$.
The idea of such linearization is that for a cold plasma model (corresponding to $n_0$), there is no effect related with the medium motion and that the corrections to first order in $\varepsilon$ are caused by a small difference in the refractive index to the cold plasma case. Limiting ourselves to this simplification, in this Section, however, we consider a more general case of the medium motion, compared to the previous sections. We assume that the four-velocity of the medium contains both $V^r$ and $V^\varphi$ components different from zero, namely
\begin{equation}
V^i = (V^0, V^r, 0, V^\varphi) \, .
\end{equation}
We assume that both components are given functions of $r$ only: $V^r=f_1(r)$, $V^\varphi=f_2(r)$. As a result, $p_0$ and $p_\varphi$ are still constants of motion. The photon frequency $\omega$ will contain both $p_r$ and $p_\varphi$ components of momentum:
\begin{equation}
\omega(r, p_r, p_\varphi) = -p_0 V^0 - p_r V^r - p_\varphi V^\varphi \, ,    
\end{equation}
or, together with $p_0 = - \omega_0$,
\begin{equation} \label{omega-general-components}
\omega(r, p_r, p_\varphi) = \omega_0 V^0(r) - p_r f_1(r) - p_\varphi f_2(r) \, .   
\end{equation}
As usual, the value of $V^0(r)$ can be found from normalization of $V_i V^i = -1$.

With the refractive index in the form (\ref{eq:refr-index-linearized}), the function $w$ defined by Eq.~(\ref{def-w}), in linear approximation in $\varepsilon$, is given by
\begin{equation}
w = - (n_0^2-1) \, \omega^2 - 2 \varepsilon n_0 n_1 \omega^2 \, .
\end{equation}
We rewrite it as
\begin{equation}
w = w_0 + 2 \varepsilon w_1 \, ,
\end{equation}
where
\begin{equation}
w_0 = \omega_p^2(r) \, , \quad  w_1 = - n_0 n_1 \omega^2 \, .
\end{equation}
In particular, $w_0$ is independent of velocities or momenta.

The definition of $w_1$ is further used when dealing with the Hamiltonian, which now reads
\begin{align} \label{eq:Hamiltonian-perturbative}
\begin{split}
\mathcal{H}(r,\omega_0,p_r,p_\varphi) &= \frac{1}{2} \left[\frac{p_r^2}{B(r)} + \frac{p_\varphi^2}{D(r)} - \frac{\omega_0^2}{A(r)} + \omega_p^2(r)\right] \\
&+ \varepsilon w_1(r,\omega_0,p_r,p_\varphi) \,.
\end{split}
\end{align}

For the calculation of the deflection angle in the medium with the Hamiltonian (\ref{eq:Hamiltonian-perturbative}) we will follow the method from the paper by L{\"a}{\"a}nemets, Hohmann, and Pfeifer \cite{Laanemets-Hohmann-Pfeifer-2022} in the following four steps.

First, we will use the fact that at the point of the closest approach  $r=R$ we have that $\dot r = 0$ and so
\begin{align}
\begin{split}
    &\frac{\partial}{\partial p_r}\mathcal{H}(R,\omega_0,p_r, p_\varphi) \\
    &= \frac{p_r}{B(R)} + \varepsilon  \frac{\partial}{\partial p_r} w_1(R,\omega_0,p_r,p_\varphi) 
    = 0\,.
\end{split}
\end{align}
This equation can be solved for $p_r(R,\omega_0, p_\varphi)$ order by order by setting
\begin{align}\label{eq:prpert1}
    p_r = p_r(R,\omega_0, p_\varphi) = 0 - \varepsilon B(R) \frac{\partial}{\partial p_r} w_1(R,\omega_0,0,p_\varphi)\,.
\end{align}
Second, since we are considering the motion of light, we use \eqref{eq:prpert1} to solve $\mathcal{H}
(R,\omega_0,p_r(R,\omega_0, p_\varphi), p_\varphi)=0$ for
\begin{align}\label{eq:pphipert}
\begin{split}
    p_\varphi 
    = p_\varphi(R,\omega_0)
    =  \bar p_\varphi + \varepsilon p_{\varphi\varepsilon}\,,
\end{split}
\end{align}
with
\begin{align}
    \bar p_\varphi = \pm \sqrt{ \frac{D(R) (\omega_0^2-A(R) \omega^2_p(R)) }{A(R)}}\,,
\end{align}
and
\begin{align}
   p_{\varphi\varepsilon} = - D(R) \frac{w_1(R, \omega_0,0,\bar{p}_\varphi)}{\bar p_\varphi(R,\omega_0)}\,.
\end{align}
to parametrize all our findings in terms of $R$ and $\omega_0$. We introduced the notations $\bar p_\varphi$ for the value of $p_\varphi$ in the limit $\varepsilon\to 0$. The symbol $p_{\varphi\varepsilon}$ denotes the first order correction induced by $w_1$.

The alternative, solving $\mathcal{H}(R,\omega_0,p_r(R,\omega_0, p_\varphi), p_\varphi)=0$ for $\omega_0 = \omega_0(R,p_\varphi)$ in order to parametrize all expressions in terms of $R$ and $p_\varphi$, will be discussed at the end of the section.

Next, third, we can solve the dispersion relation  $\mathcal{H}(r,\omega_0,p_r, p_\varphi)=0$ at any point $r$ and use \eqref{eq:pphipert} to obtain the radial momentum at any point
\begin{align}\label{eq:prOFrRomega0}
    p_r 
    &= p_r(r, R, \omega_0) = \bar p_r + \varepsilon p_{r\varepsilon}\,,
\end{align}
with zeroth order term
\begin{align}
    \bar p_r = \pm \sqrt{B(r)} \sqrt{ \frac{\omega_0^2}{A(r)} - \frac{ \bar p_\varphi^2}{D(r)}- \omega^2_p(r)  }
\end{align}
and first order correction
\begin{align}
    p_{r\varepsilon} = - \frac{B(r)}{D(r)} \frac{1}{\bar p_r} \left[ \bar p_\varphi p_{\varphi\varepsilon} + D(r) w_1(r, \omega_0,\bar p_r,\bar p_\varphi) \right] \, .
\end{align}
Fourth and final, we can use the expressions we obtained above to derive the integrand of the deflection angle.
Using \eqref{eq:prOFrRomega0} and \eqref{eq:pphipert}, we can express the integrand for the deflection angle in the variables $(r,R,\omega_0)$
\begin{align} \label{eq:defanglepert1}
    \begin{split}
    &\frac{d\varphi}{dr}(r,R,\omega_0) 
    = \frac{\dot \varphi}{\dot r} 
    = \frac{ \frac{\partial \mathcal{H} }{\partial p_\varphi} }{\frac{\partial \mathcal{H}}{\partial p_r}}
    = \frac{\bar p_\varphi}{\bar p_r}\frac{B(r)}{D(r)} \\
    &+ \varepsilon \frac{1}{\bar p_r^2} \frac{B(r)}{D(r)}
    \bigg\{
    \bar p_r \left[ p_{\varphi\varepsilon} + D(r) \tfrac{\partial w_1}{\partial p_\varphi}(r, \omega_0,\bar p_r,\bar p_\varphi) \right] \\
    &- \bar p_\varphi \left[  p_{r\varepsilon} + B(r)\tfrac{\partial w_1}{\partial p_r}(r, \omega_0,\bar p_r,\bar p_\varphi) \right]
    \bigg\} \, .
    \end{split}
\end{align}
Using the relation (\ref{omega-general-components}), we can further express the derivatives $\partial w_1/\partial p_r$ and $\partial w_1/\partial p_\varphi$ in (\ref{eq:defanglepert1}) as
\begin{equation}
\frac{\partial w_1}{\partial p_r}  = - \frac{\partial w_1}{\partial \omega}  f_1(r) \, ,  \quad \frac{\partial w_1}{\partial p_\varphi}  = - \frac{\partial w_1}{\partial \omega}  f_2(r) \, .
\end{equation}
This in principle concludes the derivation of the integrand from which the deflection angle is obtained by integration, in the same manner as in Eq.~\eqref{eq:traj-three-ex} and Eq.~\eqref{eq:angle-three-ex}.

However, for completeness we want to demonstrate how this result would look like when parametrised in terms of the angular momentum $p_\varphi$ instead of the frequency $\omega_0$.

The alternative path in the second step is to solve $\mathcal{H}(R,\omega_0,p_r(R,\omega_0, p_\varphi), p_\varphi)=0$ for
\begin{align}\label{eq:omega0pert}
\begin{split}
    \omega_0 
    &= \omega_0(R,p_\varphi) 
    = \bar \omega_0  + \varepsilon \omega_{0\varepsilon} \\
\end{split}
\end{align}
with 
\begin{align}
    \bar \omega_0 = \sqrt{ \frac{A(R) ( D(R) \omega^2_p(R)+ p_\varphi ^2) }{D(R)}}\,,
\end{align}
and
\begin{align}
    \omega_{0\varepsilon} = A(R) \frac{w_1(R,\bar{\omega}_0,0,p_\varphi)}{\bar \omega_0}\,.
\end{align}
The third step is again to use the dispersion relation $\mathcal{H}(r,\omega_0,p_r, p_\varphi)=0$ at any point $r$ and to use \eqref{eq:omega0pert} to obtain 
\begin{align}\label{eq:prOFrRpphi}
    p_r 
    &= p_r(r, R, p_\varphi) = \bar p_r + \varepsilon p_{r\varepsilon}\,,
\end{align}
with
\begin{align}
\bar p_r=\pm \sqrt{B(r)} \sqrt{ \frac{\bar \omega_0^2}{A(r)} - \frac{ p_\varphi^2}{D(r)}- \omega_p^2(r) }
\end{align}
and
\begin{align}
    p_{r\varepsilon} =  \frac{B(r)}{A(r)} \frac{1}{\bar p_r} \left[ \bar\omega_0 \omega_{0\varepsilon} - A(r) w_1(r,\bar \omega_0,\bar p_r,p_\varphi) \right] \,.
\end{align}
Now, the integrand of the deflection angle can be derived as function of $(r,R,p_\varphi)$ using \eqref{eq:prOFrRpphi} and \eqref{eq:omega0pert}
\begin{align}\label{eq:defanglepert2}
\begin{split}
    &\frac{d\varphi}{dr}(r,R,p_\varphi)
    = \frac{\dot \varphi}{\dot r} 
    = \frac{ \frac{\partial \mathcal{H} }{\partial p_\varphi} }{\frac{\partial \mathcal{H}}{\partial p_r}}
    = \frac{p_\varphi}{\bar p_r}\frac{B(r)}{D(r)} \\
    &+ \varepsilon \frac{1}{\bar p_r^2} \frac{B(r)}{D(r)}
    \bigg\{
    \bar p_r D(r) \tfrac{\partial w_1}{\partial p_\varphi}(r, \bar\omega_0,\bar p_r,p_\varphi) \\
    &- p_{\varphi} \left[ p_{r\varepsilon} + B(r) \tfrac{\partial w_1}{\partial p_r}(r, \bar\omega_0,\bar p_r,p_\varphi) \right] 
    \bigg\} \, .
\end{split}
\end{align}

Depending on the situation which we wish to describe or which information is accessible, one can now derive the deflection angle either as a function of $(R,\omega_0)$ by integrating \eqref{eq:defanglepert1} or as a function of $(R,p_\varphi)$ by integrating \eqref{eq:defanglepert2}.

Last but not least, we would like to remark that the whole linearisation procedure is valid for the Hamiltonian \eqref{eq:Hamiltonian-perturbative} with $\omega_p=0$. This then would describe light deflection in a dispersive medium that is so thin that the deviation from vacuum light propagation is very small.

\section{Conclusions}
\label{sec:conclusions}

In this paper, we have investigated gravitational deflection of light rays in a spherically symmetric spacetime filled with a moving dispersive medium characterized by its refractive index. We exploit the approach of Synge \cite{Synge-1960}, which is applicable for geometrical optics in a curved spacetime in the presence of a medium and it is widely used in the literature. The main ideas of Synge's approach and specific properties of the cold plasma case were reviewed in Section \ref{sec:cold-plasma}. Previous works on this subject have usually considered either particular case of cold nonmagnetized plasma (in which medium motion does not affect the angle of deflection) or a general medium of arbitrary refractive index, but only static (see, e.g., Tsupko \cite{Tsupko-2021}). Here, the general dispersive medium in motion has been considered.

We have studied two physically motivated scenarios of the medium motion (Fig.~\ref{fig:two-models}), for which we have written down the Hamiltonian, the equations of motion and then derived the deflection angle. 
\begin{enumerate}
    \item First, we assumed a spherically symmetric accretion of matter onto a gravitating object (see Sections \ref{sec:radial-falling} and \ref{sec:three-examples}). Consequently, only the radial spatial component of the velocity is present, and we set that it depends solely on the radial coordinate and not on time, i.e., $V^r=f(r)$, $V^\varphi=0$. This guarantees that $p_0$ is a constant of motion even in the moving medium.

    We have shown that for the refractive index of a rather general form \eqref{eq:refr-a0-a1-a2}, the deflection angle can be found analytically in a compact form given by Eq.~\eqref{eq:angle-three-ex} (Section \ref{sec:three-examples}). As particular examples, we considered three media: cold plasma case with refractive index \eqref{refr-index-plasma-r}, refractive but non-dispersive medium characterized by \eqref{refr-index-n-r}, and dispersive medium with a specific refractive index in the form \eqref{refr-index-third-example}. Comparison of the results obtained for these examples clearly shows which terms occur in the deflection angle due to the presence of a medium motion.
    \item
    As the second scenario, we have considered a rotating accretion disk in the equatorial plane, where $\varphi$-motion is present: $V^\varphi=f(r)$, $V^r=0$. This is described in Section \ref{sec:phi-motion}. In this case, we assumed that a light ray propagates in the disk plane. The resulting expression for the deflection angle is again quite compact and it is given by \eqref{eq:angle-phi-motion}. Note that in comparison to the previous case, the deflection angle has two solutions, depending on whether the light propagates in the same direction as the disk rotates or opposite.
\end{enumerate}

The main aim of this paper was to consider media beyond the cold plasma case. For cold plasma, the motion of the medium has no effect on the light deflection. This result was previously known, and we discussed that in detail in Section \ref{sec:cold-plasma}. Thus, dependence of the deflection angle on the medium velocity is relevant only for more general refractive indices. In other words, dependence of the deflection angle on the medium velocity is an indicator that the medium is \textit{not} cold plasma and it is most probably described by a more complex model. 

With this motivation, we have considered the refractive index in the form of a cold plasma with a small perturbative term of a general form, given as Eq.~\eqref{eq:refr-index-linearized}. Thus, it was possible to calculate the deflection angle perturbatively up to the first order (Section \ref{sec:general}). Then, the null term corresponds to the cold plasma case and thus does not depend on the medium velocity, while the linear term is present due to the medium velocity arising from the dependence on the photon frequency; see formulas \eqref{eq:defanglepert1} and \eqref{eq:defanglepert2}.

\section*{Acknowledgements}

The authors gratefully thank Volker Perlick for useful discussions during the work on this paper and comments about the text of the manuscript. BB and OYuT would also like to thank Ji\v{r}\'{i} Bi\v{c}\'{a}k in memoriam for his lively interest about the topic and relevant comments to particular problems.
BB is supported by UNCE24/SCI/016 grant. The work of OYuT is supported by a Humboldt Research Fellowship for experienced researchers from the Alexander von Humboldt Foundation; OYuT thanks Claus Lämmerzahl for great hospitality  at ZARM, Bremen University. CP acknowledges the financial support by the excellence cluster QuantumFrontiers of the German Research Foundation (Deutsche Forschungsgemeinschaft, DFG) under Germany's Excellence Strategy -- EXC-2123 QuantumFrontiers -- 390837967 and was funded by the Deutsche Forschungsgemeinschaft (DFG, German Research Foundation) - Project Number 420243324.

\bibliography{biblio}

\end{document}